\documentclass[aps,prl,reprint,amsmath,amssymb,showpacs]{revtex4-1}
\usepackage{graphicx}
\usepackage{float}
\usepackage{longtable}
\usepackage{slashbox}
\usepackage{subfigure}
\usepackage{lineno}
\usepackage[pdftex,  
            pdfstartview=FitH,
            CJKbookmarks=true,
            bookmarksnumbered=true,
            bookmarksopen=true,
            colorlinks, 
            pdfborder=001,   
            linkcolor=black,
            anchorcolor=blue,
            citecolor=blue,
            urlcolor=blue
            ]{hyperref}

\begin{document}
%\linenumbers
%\renewcommand\linenumberfont{\normalfont\small}
%\preprint{APS/123-QED}

\title{Experimental Demonstration of Polarization Encoding 
\\Measurement-Device-Independent Quantum Key Distribution}
%\thanks{A footnote to the article title}

\author{Zhiyuan Tang}
\email{ztang@physics.utoronto.ca}
\author{Zhongfa Liao}
\email{zhongfa.liao@mail.utoronto.ca}
\author{Feihu Xu}
\email{feihu.xu@utoronto.ca}
\author{Bing Qi}
\email{bqi@physics.utoronto.ca}
\author{Li Qian}
\email{l.qian@utoronto.ca}
\author{Hoi-Kwong Lo}
\email{hklo@comm.utoronot.ca}


\affiliation{
 Centre for Quantum Information and Quantum Control\\
 Department of Physics \& Department of Electrical and Computer Engineering\\ 
 University of  Toronto,
 Toronto, Ontario, Canada M5S 3G4
 }


\date{\today}

\begin{abstract}
We demonstrate the first implementation of polarization encoding measurement-device-independent quantum key distribution (MDI-QKD), which is immune to all detector side-channel attacks. Active phase randomization of each individual pulse is implemented to protect against attacks on imperfect sources. By optimizing the parameters in the decoy state protocol, we show that it is feasible to implement polarization encoding MDI-QKD over large optical fiber distances. A 1600-bit secure key is generated between two parties separated by 10 km of telecom fibers. Our work suggests the possibility of building a MDI-QKD network, in which complicated and expensive detection system is placed in a central node and users connected to it can perform confidential communication by preparing polarization qubits with compact and low-cost equipment. Since MDI-QKD is highly compatible with the quantum network, our work brings the realization of quantum internet one step closer.

%Moreover, our work paves the way for future %implementation of free space MDI-QKD with an untrusted %satellite. 
\end{abstract}

%\keywords{}

\maketitle

%\section{Introduction}
%\subsection{}
%\subsubsection{}

Quantum key distribution (QKD) allows two parties, normally referred to as Alice and Bob, to generate a private key even with the presence of an eavesdropper, Eve \cite{BB84,PhysRevLett.67.661}. With perfect single photon sources and single photon detectors, the security of QKD is guaranteed by quantum mechanics \cite{QKDMAYERS,*QKDLO, *QKDSHOR}. However, the aforementioned perfect devices are not available today and the security of QKD cannot be guaranteed in real life implementation. For example, attenuated coherent laser pulses are commonly used in practical QKD setups, which makes the QKD system vulnerable to the photon number splitting (PNS) attack \cite{PhysRevLett.85.1330}. Fortunately, it has been shown that the unconditional security of QKD can still be assured with phase randomized weak coherent pulses \cite{GLLP}. Furthermore, by applying decoy state techniques \cite{PhysRevLett.91.057901,*PhysRevLett.94.230504,*PhysRevLett.94.230503}, secure key rate can be dramatically increased in practical implementations \cite{PhysRevLett.96.070502,*DecoyIEEE,*PhysRevLett.98.010503,*PhysRevLett.98.010505}. Nonetheless, other imperfections in practical QKD systems still present loopholes that can be exploited by Eve to steal the secret key  \cite{QIC.7.073,*PhysRevA.78.042333,*NaturePhotonics,*NatureComm,*PhysRevLett.107.110501,*NJP.13.073024,PhysRevA.75.032314, *NJP.12.113026}. We remark that most of the identified security loopholes are due to imperfections in the detection systems \cite{QIC.7.073,*PhysRevA.78.042333,*NaturePhotonics,*NatureComm,*PhysRevLett.107.110501,*NJP.13.073024}.

Much effort has been put to build loophole-free QKD systems with practical devices. On one hand, people have been trying to build a better model to understand all the imperfections in a QKD detection system \cite{Opt.Express.20}, but it is almost impossible to guarantee that all the loopholes have been fixed. On the other hand, full device-independent QKD (DI-QKD) has been proposed to close all the loopholes due to devices' imperfections \cite{DIQKD, *PhysRevLett.98.230501}. The security of DI-QKD relies on the violation of Bell's inequality and does not require any knowledge of how practical QKD devices work. However, the demand for single photon detectors with near unity detection efficiency and the low key rate make this protocol highly impractical \cite{PhysRevLett.105.070501}.  

\begin{figure}[b]
  \centering
  \includegraphics[width=0.25\textwidth]{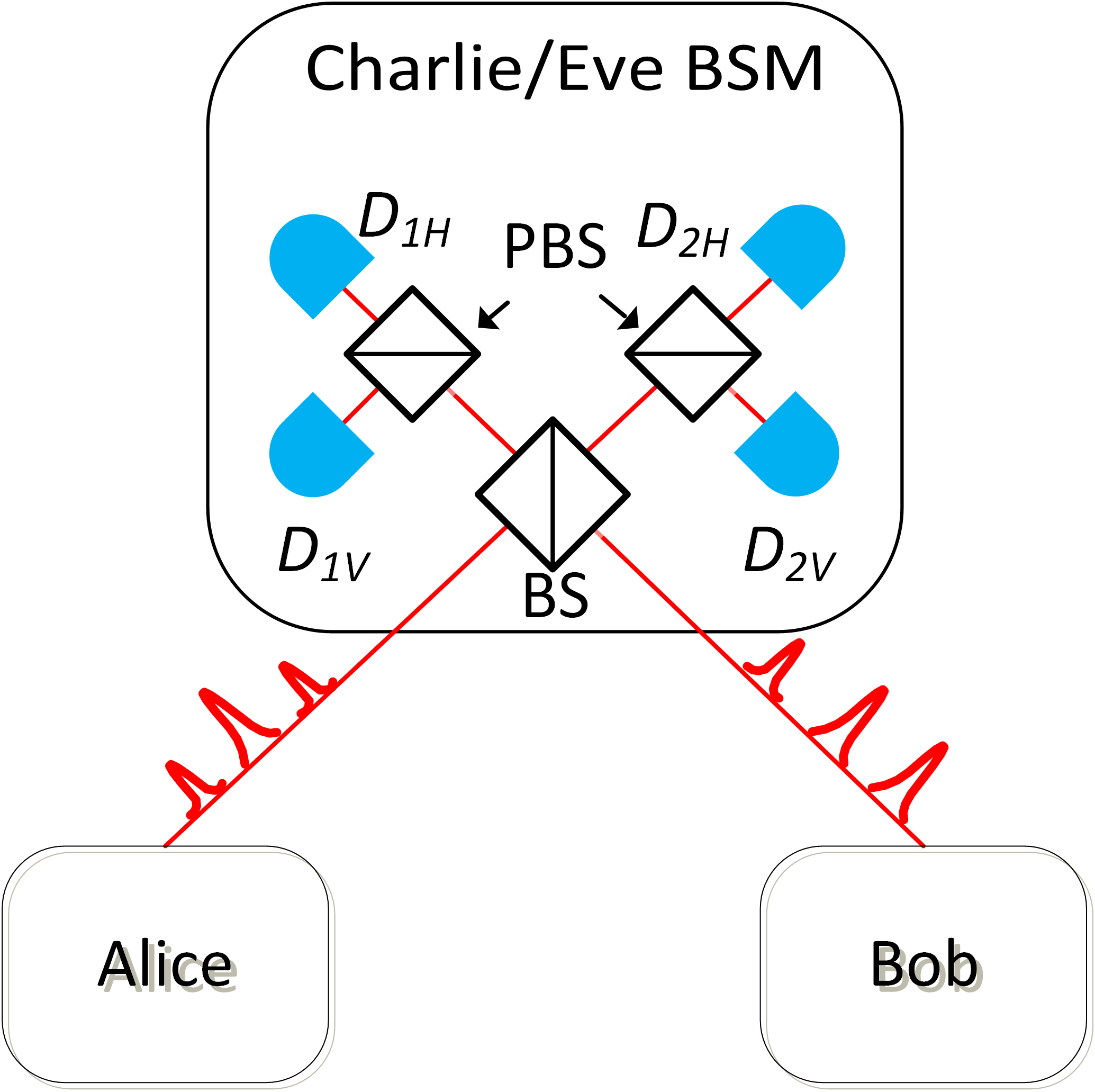}
  \caption{\label{graph}(Color online). A typical MDI-QKD setup. Alice and Bob send out polarization encoded weak coherent pulses (with decoy states) to Charlie/Eve, who is supposed to perform partial Bell state measurements with a beam splitter (BS), polarizing beam splitters (PBS), and single photon detectors. }
 \end{figure}

Fortunately, measurement-device-independent QKD (MDI-QKD), which removes all loopholes in detectors \cite{PhysRevLett.108.130503}, has been proposed as an alternative solution.  Although in MDI-QKD, the assumption of almost-perfect state preparation cannot be removed, finite basis-dependent flaw can be tolerated and taken care of \cite{PhysRevA.85.042307} using the quantum coin idea \cite{GLLP}.  

A typical (polarization encoding) MDI-QKD is illustrated in FIG. \ref{graph}. Alice and Bob independently prepare weak coherent pulses (with decoy states) in one of the four BB84 polarization states \cite{BB84}, and send them to an untrusted third party (UTP) Charlie, who can be an eavesdropper (Eve). Charlie/Eve is supposed to perform Bell state measurements (BSMs) on the incoming pulses and to publicly broadcast the BSM results to Alice and Bob. A partial BSM can be realized with linear optics and two of the four Bell states can be registered: a coincidence between detectors $D_{1H}$ and $D_{1V}$, or $D_{2H}$ and $D_{2V}$, indicates a successful projection into the triplet state $|\psi^+\rangle=1/\sqrt{2}(|HV\rangle+|VH\rangle)$, while a coincidence between detectors $D_{1H}$ and $D_{2V}$, or $D_{1V}$ and $D_{2H}$, indicates a successful projection into the singlet state $|\psi^-\rangle=1/\sqrt{2}(|HV\rangle-|VH\rangle)$, where $H$ and $V$ represent the horizontal and vertical polarization states, respectively. Alice and Bob then reveal their choices of bases over an authenticated channel and discard coincidence events where they use different bases to generate a sifted key.  A secret key can be generated after error correction and privacy amplification. In the asymptotic limit of an infinitely long key, the key rate $R$ is given by \cite{PhysRevLett.108.130503}
\begin{equation}
\label{keyrate}
R\geq q\{p_{11}Y_{11}^{Z}[1-H(e_{11}^{X})]-Q_{\mu\mu}^{Z}f(E_{\mu\mu}^{Z})H(E_{\mu\mu}^{Z})\},
\end{equation}
where $q$ is the proportion of pulses where both Alice and Bob send out signal states in the rectilinear (Z) basis, $\mu$ is the average photon number of the signal state, $p_{11}=\mu^2e^{-2\mu}$ is the conditional probability that both Alice and Bob send out single photon states given that both of them send pulses in the signal state, $Y_{11}^{Z}$ is the yield of single photon states in the rectilinear basis, $e_{11}^{X}$ is the quantum bit error rate (QBER) of single photon states in the diagonal (X) basis, $Q_{\mu\mu}^{Z}$ and $E_{\mu\mu}^{Z}$  are the gain and QBER of the signal state in the rectilinear basis,respectively, $f(E_{\mu\mu}^{Z})> 1$ is the inefficiency function of error correction, and $H(x)=-xlog_2(x)-(1-x)log_2(1-x)$ is the binary Shannon entropy. Here $Q_{\mu\mu}^Z$ and $E_{\mu\mu}^Z$ can be directly measured from the experiment, and a lower bound of $Y_{11}^{Z}$ and an upper bound of $e_{11}^{X}$ are to be estimated using the decoy state technique. 

\begin{figure*}
  \centering
  \subfigure[]{\label{setup}\includegraphics[width=0.75\textwidth]{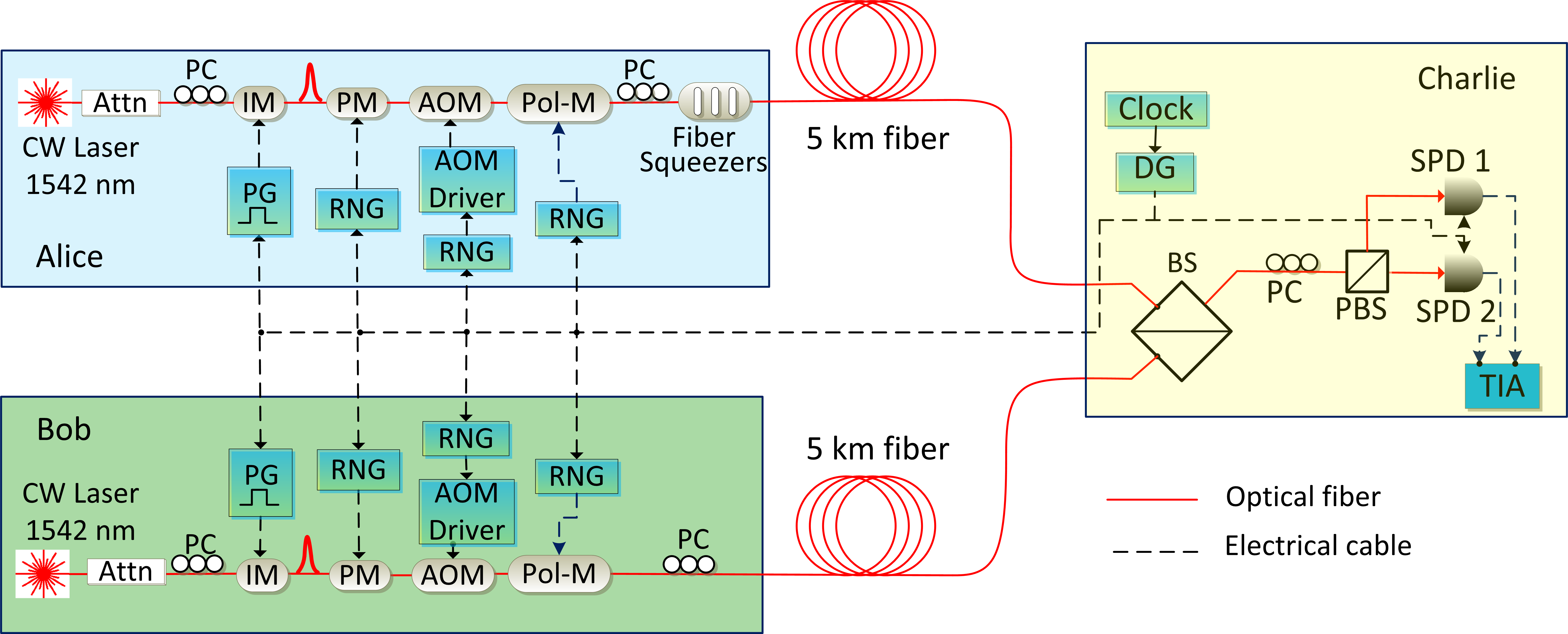}}                
  \subfigure[]{\label{polm}\includegraphics[width=0.24\textwidth]{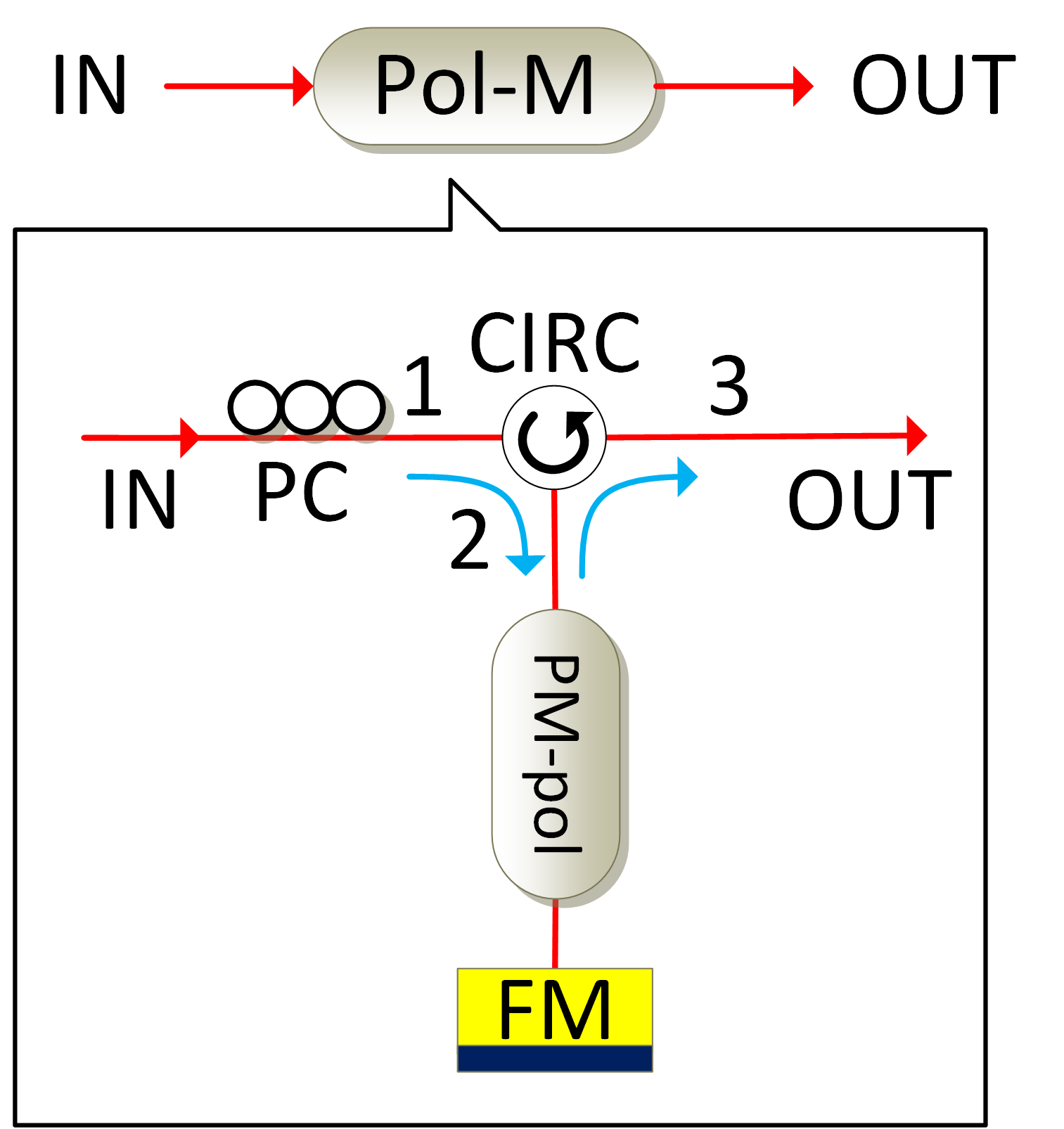}}
  \caption{(Color online). (a) Experimental setup of polarization encoding MDI-QKD: Attn, optical attenuator; IM, intensity modulator, PM, phase modulator; AOM, acousto-optic modulator; Pol-M, polarization modulator; PC, polarization controller; PG, electrical pulse generator; RNG, random number generator; DG, delay generator; BS, beam splitter; PBS, polarizing beam splitter; SPD, single photon detector; TIA, time interval analyser. (b) Schematic of the polarization modulator: CIRC, optical circulator; PM-pol, phase modulator; FM, Faraday mirror; PC, polarization controller.}
\end{figure*}

The idea of MDI-QKD is indeed built on the time-reversed EPR-based QKD protocol \cite{PhysRevA.54.2651,*Algorithmica.34.340}. Intuitively, the security of MDI-QKD relies on the fact that Charlie/Eve is post-selecting entanglement between Alice and Bob, who can verify such post-selected entanglement via authenticated public discussion of their polarization data.

Since the detection system can be placed in an untrusted third party in MDI-QKD, the need for certifying detectors, which has been a major effort in the standardization of QKD \cite{NJP.11.055051}, is completely removed. This also means that Alice and Bob can freely outsource the manufacturing process to untrusted manufacturers. Furthermore, MDI-QKD can be implemented in a network setting: Charlie/Eve, who possesses expensive single photon detectors, serves as a service centre, and multiple users can connect to this centre and share the detection system using time or wavelength multiplexing techniques, thus leading to a significant cost reduction in building a MDI-QKD network. In addition, the transmission distance of MDI-QKD can be further extended via EPR pairs \cite{1306.5814} and other quantum relay and quantum repeater solutions \cite{1306.3095}, thus highlighting its compatibility with the future quantum internet \cite{Nature.453.1023}.

Various experimental attempts of MDI-QKD with time-bin encoding \cite{1304.2463, 1209.6178} and polarization encoding \cite{1207.6345} have been reported . We remark that in \cite{1304.2463, 1207.6345}, only Bell state measurements with different BB84 states and intensities were conducted, and in fact no real MDI-QKD, which requires that both Alice and Bob randomly and independently modulate their qubits' states and intensity levels, was performed. Additionally, phase randomization of weak coherent pulses \cite{APL.90.4.044106}, a crucial assumption in security proofs of decoy state QKD \cite{PhysRevLett.94.230504,PhysRevLett.94.230503}, was neglected in their implementation, leaving the system vulnerable to attacks on the imperfect weak coherent sources. In fact, an unambiguous-state-discrimination (USD) attack on a decoy state QKD system without phase randomization has been reported recently \cite{1304.2541}. A time-bin phase encoding MDI-QKD was performed in \cite{1209.6178}. To defend against the USD attack, they turned the lasers on and off for each pulse and assumed that the phase of each individual pulse is randomized. However, intensity levels and probability distribution of the signal and decoy states were not optimized. 
 
In this paper, we report the first experimental demonstration of polarization encoding MDI-QKD over 10 km of optical fibers. While phase and time-bin encoding MDI-QKD has the advantage of easier polarization management in optical fiber, it requires expensive and bulky equipment for phase stabilization in interferometers at the users' (Alice and Bob) sites. On the other hand, polarization qubits can be easily prepared by the users using standard optoelectronic devices, which can be miniaturized using state-of-the-art micro-fabrication processes at a low cost \cite{1305.0305}. Those more expensive polarization stabilization systems, if required, can be placed with other expensive equipments (e.g., detectors) at the service centre. This makes polarization encoding more favourable in a network setting. Moreover, there is an increasing interest in implementing QKD in free space, particularly ground-to-satellite QKD \cite{PhysRevA.84.062326,*NP.7.382,*NP.7.387}, in which polarization is a preferred encoding scheme. Our work thus paves the way for future implementation of ground-to-satellite MDI-QKD, in which measurement devices can be placed in an untrusted satellite.

We implement MDI-QKD with two decoy states \cite{PhysRevA.72.012326} over 10 km of telecommunication fibers. We perform a numerical simulation to optimize the performance \cite{1305.6965}: average photon numbers are chosen to be $\mu=0.3$ for the signal state, $\nu=0.1$ and $\omega=0.01$ for the two decoy states \footnote{Ideally the decoy state $\omega$ should be set to vacuum in order to optimize the performance. Since the extinction ratios of our acousto-optic modulators are finite, it is impossible to generate perfect vacuum state in practice. A low (but non-zero) $\omega$ (e.g., $\omega=0.001$) would give a key rate close to that when $\omega$ is vacuum. However, given the data size in our experiment, it is difficult to measure the gain $Q_{\omega\omega}^{X,Z}$ and the error rate $E_{\omega\omega}^{X,Z}$ for such a small $\omega$. Although the choice of $\omega=0.01$ slightly compromises the performance, it is more practical for our system.}; the ratio of numbers of pulses sent out with intensities $\mu$, $\nu$, and $\omega$ is set to be $4:9:7$. Active phase randomization is implemented to defend against attacks on the imperfect weak coherent sources.

Figure \ref{setup} shows the schematic of our polarization encoding MDI-QKD experiment. Each of Alice and Bob possesses a CW frequency-locked laser (Clarity-NLL-1542-HP, wavelength $\sim1542$ nm). The laser light is attenuated and modulated by a $\mathrm{LiNbO}_3$  based intensity modulator (IM) to generate weak coherent pulses at a repetition rate of 500 KHz. Phases of pulses are uniformly randomized in the range of $[0, 2\pi]$ by a phase modulator (PM). To implement the decoy state protocol, intensities of the pulses  are randomly modulated by an acousto-optic modulator (AOM) between the signal and decoy states. 

Key bits are encoded into polarization states of the weak coherent pulses by a polarization modulator (Pol-M), whose design is proposed in \cite{NJP.11.095001}. The schematic of the polarization modulator, consisting of an optical circulator, a phase modulator (labelled as PM-pol), and a Faraday mirror, is shown in FIG. \ref{polm}. Optical pulses are launched via the optical circulator into the phase modulator with polarization at $45^{\circ}$ from the optical axis of the phase modulator's waveguide. By modulating the relative phase between the two principal modes of the waveguide, four BB84 polarization states can be generated.  Polarization mode dispersion and temperature-induced variation of polarization states inside the Pol-M setup can be compensated when pulses are reflected by a Faraday mirror with a $90^{\circ}$ rotation in polarization, thus stable polarization modulation can be achieved.

Alice and Bob need to ensure that they have a shared reference frame of polarization. Their rectilinear bases ($H$ and $V$) are first aligned manually using fiber polarization controllers (PCs). Alice's $H$ polarization state is also aligned to either the fast or slow axis of one of the fiber squeezers (driven by piezoelectric actuators) in an electrical polarization controller (General Photonics PolaRITE III, in Alice's setup). A DC voltage is then applied on this squeezer to change the phase retardation between the components with polarizations along the fast and slow axes. This corresponds to a unitary transformation in which Alice's polarization states in the rectilinear basis remain unchanged, while polarization states in the diagonal basis ($|\pm\rangle=1/\sqrt{2}(|H\rangle\pm |V\rangle)$) are rotated about the $H$-$V$ axis on the Poincar\'e sphere.  The voltage is properly adjusted such that Alice's diagonal basis is aligned to Bob's. The misalignment is around 1\% in our experiment.

 All the modulators (PMs, AOMs, and Pol-Ms) are independently driven by random number generators (function generators with pre-stored random numbers generated by a quantum random number generator \cite{Opt.Lett.35.312,*Opt.Express.20.12366,*CLEO.QRNG}). An electrical delay generator (DG) located in Charlie's setup synchronizes all the RNGs and the electrical pulse generators (PGs) driving the IMs.

\begin{table} %add [H] placement to break table across pages
\caption{\label{Gains}Experimental values of gains $Q_{I_{A}I_{B}}^{W}$ ($\times10^{-4}$) with intensities $I_{A}$ and $I_{B}$ in basis $W\in\{X,Z\}$.}
\begin{ruledtabular}
\begin{tabular}{c|ccc|ccc}
   \quad & \multicolumn{3}{c}{rectilinear (Z) basis}& \multicolumn{3}{c}{diagonal (X) basis} \\ 
   \hline
    \backslashbox{$I_{B}$}{$I_{A}$} & $\mu$ & $\nu$ & $\omega$ & $\mu$ & $\nu$ & $\omega$\\
    \hline
    $\mu$ & 0.466 & 0.1597 & 0.0225 & 0.903 & 0.410 & 0.254 \\
    $\nu$ & 0.1550 & 0.0531 & 0.0070 & 0.397 & 0.1015 & 0.0312 \\
    $\omega$ & 0.0214 & 0.0067 & 0.0009 & 0.246 & 0.0317 & 0.0014 \\
\end{tabular}
\end{ruledtabular}
\end{table}

\begin{table} %add [H] placement to break table across pages
\caption{\label{QBERs}Experimental values of QBERs $E_{I_{A}I_{B}}^{W}$ with intensities $I_{A}$ and $I_{B}$ in basis $W\in\{X,Z\}$.}
\begin{ruledtabular}
\begin{tabular}{c|ccc|ccc}
 \quad & \multicolumn{3}{c}{rectilinear (Z) basis}&\multicolumn{3}{c}{diagonal (X) basis} \\ 
   \hline
    \backslashbox{$I_{B}$}{$I_{A}$} & $\mu$ & $\nu$ & $\omega$ & $\mu$ & $\nu$ & $\omega$\\
    \hline
    $\mu$ & 0.0178 & 0.0320 & 0.167 & 0.262 & 0.326 & 0.465 \\
    $\nu$ & 0.0306 & 0.0402 & 0.161 & 0.322 & 0.261 & 0.431 \\
    $\omega$ & 0.156 & 0.157 & 0.23 & 0.469 & 0.430 & 0.32 \\
\end{tabular}
\end{ruledtabular}
\end{table}

In this experiment, it is critical to assure that the weak coherent pulses independently prepared by Alice and Bob are indistinguishable at Charlie's beam splitter in terms of spectrum and arrival time. The wavelengths of the lasers used by Alice and Bob are independently locked to one molecular absorption line of a gas cell (integrated in the laser by the manufacturer) at around 1542.38 nm. This guarantees the frequency difference between Alice and Bob's lasers is within 10 MHz, while the temporal width of the pulse is about 1 ns (FWHM), corresponding to a bandwidth of about 1 GHz. The arrival time of the pulses can be independently controlled by the DG with a resolution of 50 ps, and the timing jitter of the electronic devices is about 100 ps (RMS). Therefore, we can guarantee that the two independently prepared pulses have sufficient overlap in both time and spectrum.

 Alice and Bob send their pulses through a 5 km fiber spool to Charlie , who performs Bell state measurements on the incoming pulses. Charlie's measurement setup consists of a 50:50 beam splitter (BS), a polarizing beam splitter(PBS), and two commercial InGaAs/InP single photon detectors (SPDs, detection efficiency $\sim10\%$, dark count rate $\sim5\times10^{-5}$). Due to the limited number of available detectors, we choose to detect photons at the outputs of one PBS only. A coincidence between the two detectors (defined as when both SPDs click within 10 ns, measured by a time-interval analyser, TIA) in this setup, as discussed above, corresponds to a successful projection into the triplet state  $|\psi^+\rangle$.   

The experiment runs for about 94 hours \footnote{Polarizations in our system can remain stable for about one hour without active polarization feedback control. The QKD experiment is stopped every hour for polarization realignment. The realignment process can be automated by adding a polarization compensation system to the setup in the future.} and a total number of $N=1.69\times 10^{11}$ pulses are sent out. The gains and QBERs with different intensities in the rectilinear and diagonal bases can be measured from the sifted key and the results are listed in Tables \ref{Gains} and \ref{QBERs}, respectively. 

To extract a secure key, Alice and Bob need to estimate a lower bound of $Y_{11}^{Z}$ (denoted as $Y_{11}^{Z,L}$) and an upper bound of $e_{11}^{X}$ (denoted as $e_{11}^{X,U}$) from the finite decoy-state protocol. Here we use an analytical method with two decoy states for our finite decoy-state analysis. For an infinitely long key, $Y_{11}^{Z,L}$ and $e_{11}^{X}$ are given by~\cite{1305.6965}
\begin{widetext}
\begin{equation} 
\label{Y11L}
\begin{split}
Y_{11}^{Z,L}=&\dfrac{1}{(\mu-\nu)^{2}(\nu-\omega)^{2}(\mu-\nu)}\\
&\times [(\mu^2-\omega^2)(\mu-\omega)(e^{2\nu}Q^{Z}_{\nu\nu}+e^{2\omega}Q^{Z}_{\omega\omega}-e^{\nu+\omega}Q^{Z}_{\nu\omega}-e^{\omega+\nu}Q^{Z}_{\omega\nu})\\
&-(\nu^2-\omega^2)(\nu-\omega)(e^{2\mu}Q^{Z}_{\mu\mu}+e^{2\omega}Q^{Z}_{\omega\omega}-e^{\mu+\omega}Q^{Z}_{\mu\omega}-e^{\omega+\mu}Q^{Z}_{\omega\mu})],
\end{split}
\end{equation}
\begin{equation}
\label{e11U}
\begin{split}
e_{11}^{X,U}=\dfrac{e^{2\nu}Q^{X}_{\nu\nu}E^{X}_{\nu\nu}
+ e^{2\omega}Q^{X}_{\omega\omega}E^{X}_{\omega\omega}
-e^{\nu+\omega}Q^{X}_{\nu\omega}E^{X}_{\nu\omega}
-e^{\omega+\nu}Q^{X}_{\omega\nu,U}E^{X}_{\omega\nu}  }{(\nu-\omega)^2Y^{X,L}_{11}},
\end{split}
\end{equation}
\end{widetext}
where $Y^{X,L}_{11}$ in Eq. (\ref{e11U}), a lower bound of the gain of single photon states in the diagonal basis, can be estimated using Eq.~(\ref{Y11L}), with the superscripts $Z$ replaced by $X$. 

Since the data size in our experiment is finite, statistical fluctuations of the experimental parameters $Q_{I_{A}I_{B}}^{X,Z}$ and $E_{I_{A}I_{B}}^{X,Z}$ need to be taken into account when estimating $Y_{11}^{Z,L}$ and $e_{11}^{X,U}$ . We use the method proposed in \cite{PhysRevA.86.052305} and assume a secure bound of $\epsilon$=$1\times10^{-3}$ (three standard deviations) \footnote{We also try to perform a more rigorous finite key analysis proposed in \cite{1307.1081}, in which MDI-QKD can be made to satisfy the definition of composable security. We find that with our experimental parameters and a security bound of $\epsilon$=$1\times10^{-7}$, it requires a minimal number of $\sim10^{12}$ pulses , equipvalent to 600 hours continuous experiment due to the low speed in our system. We emphasize that our experimental data indicates it is possible to achieve composable security with our system if the experiment is run for a longer time.}. We find $Y_{11}^{Z,L}$=$4.1\times10^{-4}$ and $e_{11}^{X,U}$=15.1\%. We can estimate a lower bound of the secure key rate  $R^{L}=9.8\times10^{-9}$ using Eq. (\ref{keyrate}) and the parameters summarized in Table \ref{para_key}. Therefore, a secure key of length $L=1600$ bits can be generated between Alice and Bob.

\begin{table}
\caption{\label{para_key}Parameters used to estimate a lower bound of the key rate $R^{L}$.}
\begin{ruledtabular}
\begin{tabular}{ccccccc}
   
    $q$ & $p_{11}$ & $Y_{11}^{Z,L}$ & $e_{11}^{X,U}$ & $Q_{\mu\mu}^{Z}$ & $E_{\mu\mu}^{Z}$ & $f$ \\
     \hline
    0.011 & $0.0494$ & $4.1\times10^{-4}$ & 0.151 & $4.66\times10^{-5}$ & 0.0178 & 1.16 \\

\end{tabular}
\end{ruledtabular}
\end{table}

Compared to the conventional BB84 QKD protocol, the key rate of MDI-QKD is lower due to the fact that this novel protocol relies on coincidence (rather than single) detection events to generate key bits. This can be circumvented by using state-of-the-art superconducting single photon detectors (SSPDs) with detection efficiency over 90\% \cite{NP.7.210} at the telecom wavelengths, or alternatively, silicon detectors with 50\% detection efficiency at around 800 nm \cite{1208.4205} for free space implementation. Polarization feedback control system \cite{1207.6345} can be incorporated into a MDI-QKD system to further lower the QBERs, thus reducing the cost in error correction and privacy amplification. The key generation speed of MDI-QKD can also be substantially increased if implemented at a high repetition rate ($>$ 1 GHz), which is feasible with current technology \cite{OptExp.13.003015,*OptExp.16.003015}.

In summary, we have demonstrated the first polarization encoding MDI-QKD experiment over 10 km of optical fibers, with active phase randomization implemented to defeat attacks on imperfect sources. Our work shows that, with commercial off-the-shelf optoelectronic devices, it is feasible to build a QKD system immune to detector side-channel attack. In particular, the practicability of polarization encoding MDI-QKD indicates the potential to build a detector side-channel free QKD network, in which users only need to possess handy hardware to prepare polarization qubits. Our work can also be extended to free space polarization encoding MDI-QKD with an untrusted satellite in the future. 

\begin{acknowledgments}
We thank W. Cui and M. Curty for enlightening discussions, and H. Xu for his assistance in the experiment. Financial supports from NSERC Discovery Grant, NSERC RTI Grant, and Canada Research Chairs Program are gratefully acknowledged.  
\end{acknowledgments}

\subsection{Appendix A: Polarization alignment in the experiment}

\begin{figure}
  \centering
  \subfigure[]
  {\label{squeezers}
    \includegraphics[width=0.3\textwidth]{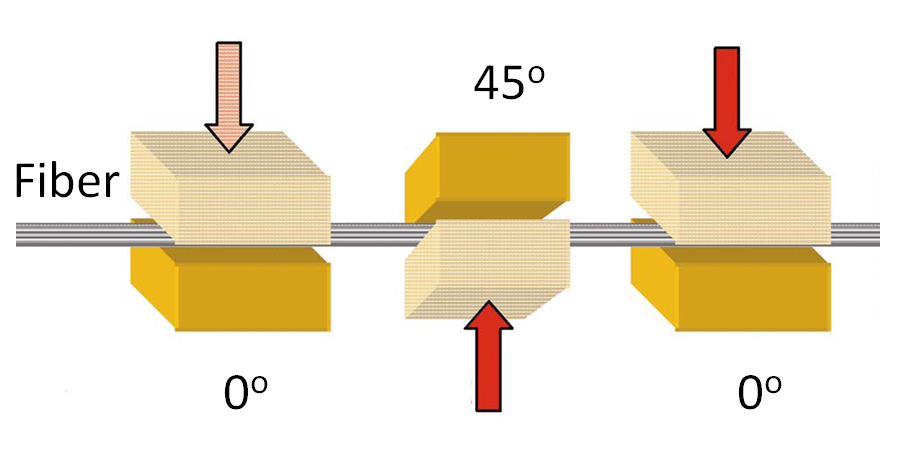}}
   \subfigure[]
   {\label{poincare}
    \includegraphics[width=0.3\textwidth]{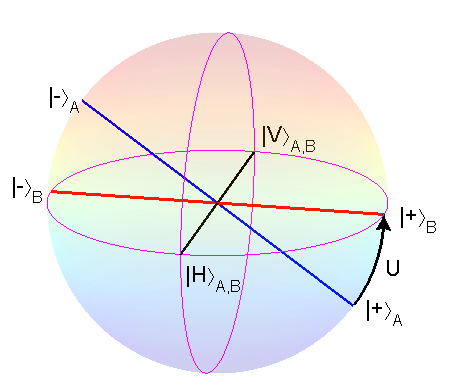}}
    \caption{ (Color online). (a)Schematic of the electrical polarization controller. Figure courtesy of General Photonics Corp. (b)Geometrical representation of  the alignment of Alice and Bob's reference frames on the Poincar\'e sphere.}
\end{figure}

In MDI-QKD, Alice and Bob independently prepare polarization qubits in their setup. For successful Bell state measurement, it requires that Alice and Bob have a shared reference frame of polarization. This is achieved by applying unitary transformations on their polarization states using fiber based polarization controllers (PCs) and an electrical polarization controller (in Alice's setup only). Figure \ref{squeezers} shows the schematic of the electrical polarization controller, which consists of three fiber squeezers (driven by piezoelectric actuators) oriented at $0^{\circ}$, $45^{\circ}$, $0^{\circ}$, respectively. Alice and Bob first align their polarization states in the rectilinear basis ($H$ and $V$) to the polarizing axes of the polarizing beam splitter (PBS) in Charlie's setup. This can be done by letting Alice/Bob send horizontally polarized pulses and adjusting PCs to minimize photon detection rates (measured by a single photon detector) at one of the outputs of the PBS. Therefore they have a shared rectilinear basis as shown in Fig. \ref{poincare}, where Alice's $H$ and $V$ states overlap with those of Bob's on the Poincar\'e sphere.

Alice's horizontal polarization state is also aligned to either the fast or slow axis of the first (from the left) fiber squeezer in the electrical polarization controller. A voltage is applied on this squeezer, and the induced pressure leads to a change in the phase retardation between the slow and fast axes. This is equivalent to a unitary transformation $U$ of Alice's diagonal basis ($|\pm\rangle=1/\sqrt{2}(|H\rangle\pm|V\rangle)$) about the $H$-$V$ axis on the Poincar\'e sphere, as illustrated in Fig. \ref{poincare}. The voltage is adjusted until Alice's diagonal basis is aligned to Bob's.  We note that this process does not disturb the alignment in the rectilinear basis done in previous step.

\end{document}